# Four-legged starfish-shaped Cooper pairs with ultrashort antinodal length scales in cuprate superconductors

Haoxiang Li[1*], Xiaoqing Zhou[1], Stephen Parham[1], Kyle N. Gordon[1], R. D. Zhong[2], J. Schneeloch[2], G. D. Gu[2], Y. Huang[3], H. Berger[4], G. B. Arnold[1], D. S. Dessau[1,5*]

[1]Department of Physics, University of Colorado Boulder, Boulder, CO 80309, USA
[2] Condensed Matter Physics and Materials Science Department, Brookhaven National Laboratory, Upton, New York 11973, USA
[3] Institute of Physics, University of Amsterdam, Amsterdam, Netherlands
[4]Départment de Physique, Ecole Polytechnique Fédérale de Lausanne, CH-1015 Lausanne, Switzerland
[5.] Center for Experiments on Quantum Materials, University of Colorado Boulder, Boulder, CO 80309, USA

**Cooper pairs of mutually attracting electrons form the building blocks of superconductivity. Thirty years after the discovery of high-temperature superconductivity in cuprates, many details of the pairs remain unknown, including their size and shape. Here we apply brand new ARPES-based methods that allow us to reconstruct the shape and size of the pairs in $Bi_2Sr_2CaCu_2O_{8+\delta}$. The pairs are seen to form a characteristic "starfish" shape that is very long (>50Å) in the near-nodal direction but extremely short (~4.5Å) in the antinodal (Cu-O) direction. We find that this ultrashort antinodal length scale, which is of order a lattice constant, is approximately constant over a wide range of doping levels even as many other parameters including the pairing strength change. This suggests that this new length scale, along with the pair shape, is one of the most fundamental characteristics of the pairs. Further, the shape and ultrashort length scale should make the pairs create or intertwine with variations in charge and pair density, center on various types of lattice positions, and potentially explain aspects of the nematic order in these materials.**

The mechanism of high-temperature superconductivity in cuprates remains one of the defining problems in condensed matter physics. By exploiting recent developments that greatly improve the quantitative accuracy of angle-resolved photoemission spectroscopy (ARPES), especially in its ability to extract the self-energies [1], we are able to directly measure all key electronic parameters and pairing interactions as a function of angle. This gives us the ability, for the first time, to reconstruct the angular-dependent pairing length scales in a superconductor (**Fig. 1**). In contrast to most previous concepts that hold that the pairs in the cuprates are round as in conventional superconductors [2], we find that the pair length scales have the shape of a 4-legged starfish. The starfish-shaped pairing lengths have very long (>50Å) arms near the node (45° to the Cu-O bonds) and a very short body of about 4.5Å along the Cu-O bonds. Previous results only indicated the overall average superconducting coherence length of the pairs to be around 20Å – a properly weighted averaging of our result returns a similar value (see supplementary information S1). As a function of doping, our data shows the surprising finding that the antinodal pair size remains approximately constant and is pinned near the lattice parameter of 3.8Å – a result that further indicates that this short length scale is not accidental but rather is likely a defining or driving characteristic of the pairing interaction.

For conventional superconductors [3], where the pair wave functions are well-known, the superconducting pair size can be defined as the expectation value of the root-mean-square radius, in either real-space or k-space, as [4,5,6]

$$\xi_{pair} = \sqrt{\frac{\int dr |\psi(r)|^2 r^2}{\int dr |\psi(r)|^2}} = \sqrt{\frac{\int dk |\partial_k \psi(k)|^2}{\int dk |\psi(k)|^2}} \quad (1)$$

where $\psi(r)$ is the Cooper-pair wave function in real space, $\partial_k$ is the gradient operator in momentum space (**k** space), and $\psi(\mathbf{k})$ is the Fourier transform of $\psi(r)$ to **k** space [7,8]. The wavefunction is $\psi(\mathbf{k})=\Delta(\mathbf{k})/E_\mathbf{k}$ with $E_\mathbf{k} = \sqrt{\varepsilon_\mathbf{k}^2 + \Delta(\mathbf{k})^2}$ [7,8], where $\varepsilon_\mathbf{k}$ is the non-gapped dispersion.

The expression of the pair size includes multiple terms (see all terms including their derivation in supplementary information S2), but the dominant term of the pair size is written as:

$$\xi_{pair}(\theta) \approx \frac{\sqrt{\int dk_\perp \frac{\varepsilon_k^2 \Delta_\theta^2}{E_k^6}(\partial_\perp \varepsilon_k)^2}}{\sqrt{\int dk_\perp \frac{\Delta_\theta^2}{E_k^2}}} \quad (2)$$

where θ is the Fermi surface angle, $E_\mathbf{k} = \sqrt{\varepsilon_\mathbf{k}^2 + \Delta_\theta^2}$, and the k-integrals and the gradients are taken in the direction perpendicular to the Fermi surface (see the justification for the dominant term in supplementary information S2 and Fig. S3). The important point is that all parameters necessary to compute Eqn. 2 are now directly accessible from ARPES experiments, especially after the recent introduction of a reliable method for the 2-dimensional analysis of ARPES data [1], which delivers consistent and accurate results for the band energy $\varepsilon_k$, the pairing gaps $\Delta_\theta$, and especially the single-particle self-energies $\Sigma'(k, \omega, T)$ and $\Sigma''(k, \omega, T)$ and the related renormalization parameter $Z(k, \omega, T)$ enter into the evaluation of $\varepsilon_k$ and $E_k$. Utilizing previously published low-temperature ARPES data [1] from a $T_C$=85K underdoped sample we extract the pair size as a function of angle, as plotted in **Fig. 1a** and **1b**. It is seen that the pair length scales have a shape reminiscent of a four-armed starfish, with the arms extending to very large distances near the nodes, with a compact body of 4.5 Å in the antinodal direction. As the gap goes to zero at the node, the pair size concept is more complicated - thus, in **Fig. 1a** and **1b**, we omitted the exact nodal direction indicated by the blank in the figure.

The above result is a natural consequence of the d-wave symmetry of the pairing interaction strengths and can be compared to the more commonly discussed Pippard coherence length [9] that is the length scale over which the superconducting state recovers when it is destroyed locally. For a BCS superconductor, the coherence length is given by

$$\xi_{CL}(\theta) = \frac{\hbar v_F}{\pi \Delta_\theta} \tag{3}$$

where $v_F$ is the Fermi velocity and $\Delta$ the superconducting gap (pairing strength) [10]. The large gap in the antinodal direction therefore naturally corresponds to the smaller length scale of the starfish body, while the very small gap near the nodes corresponds to the very long arms. Other d-wave superconductors such as some heavy fermion superconductors [11] will likely also have this shape. Fig. S1 and Fig. S2 shows our ARPES-based result of the coherence length. While previous measurements only indicated the overall average superconducting coherence length of the pairs to be around 20Å [12,13,14,15], our result, when properly weighted by an average around the zone, returns a similar value (see detailed discussion in supplementary information S1). Although previous Scanning Tunneling Spectroscopy studies only provide a single value of the pair size [15,16], they also find that the maximum gap value reflecting the antinodal gap is quite inhomogeneous on the length scale of ~15 Å, whereas the nodal gap feature is rather homogeneous, or in other words, the node has a much longer length scale. This observation suggests the existence of the anisotropic pair size in the real space. Moreover, previous STM studies [17,18] have shown some four-fold symmetric patterns generated by different impurities. In particular, the magnetic Ni impurities in ref [18] show a four-fold pattern (on the occupied side) where the separation between lobes along the antinodal direction is comparable to the Cu-Cu separation in the lattice, which is a potential connection to the ultra-short antinodal pair size. However, how these different impurity patterns compare to the shape of the Cooper pairs is unclear, and warrants further investigation.

In addition to the UD85K sample of **Fig. 1a** and **1b,** we present the antinodal pair size from 5 other superstructure-free $(Bi,Pb)_2Sr_2CaCu_2O_{8+\delta}$ samples and one $Bi_2Sr_2CaCu_2O_{8+\delta}$ sample with optimal doping (**Fig. 1c**), with the 7 samples having dopings that span the underdoped and overdoped regimes. The antinodal pair sizes are extremely short and doping independent. These antinodal pair sizes are extracted from the spectral data (**Fig. 2a-2g**) from all samples with temperature well below $T_C$ for the antinodal region of the Brillouin zone where the gap is largest (red line in **Fig. 2o**). The photon energy has been chosen so as to emphasize the antibonding band of the bilayer-split bands [19,20] – a comment about the bonding band will come later. 2D fits to these data following the procedure of ref [1] are shown in **Fig. 2h-2n**. **Fig. 2p** and **2q** also show the $k_F$ EDCs from each set of experimental data and fits [21].

The individual parameters extracted from the fits of **Fig. 2** are plotted as a function of doping in **Fig. 3**. With these individual parameters, we are then able to write the antinodal pair wavefunction and thus extract the low temperature antinodal pair size for each doping level using Eqn. 2. The results of these calculations for a wide range of doping levels are plotted in **Fig. 1c**, showing that the antinodal pair size is essentially unaffected by doping level, even though the various constituent parameters of **Fig. 3a-d** vary significantly with doping. For one doping level, we have separately extracted similar data on the bonding band of the bilayer split bands. This band is farther from the Fermi energy ($E_{BB}$ and $k_F$ are larger than the ones for the antibonding band), and though it has the same gap size as the antibonding counterpart we find that the antinodal pair size is significantly larger for the bonding band (see supplementary information S4 and Fig. S5, Fig. S6).

The most surprising aspect of these results is the extremely short and universal antinodal pair size of order 4.5Å. This is the case even though the individual parameters determining this quantity (Eqn. 2) all vary with doping (**Fig. 3**), in some cases strongly. We therefore argue that this constant ultrashort length scale, which is close to the lattice parameter of 3.8 Å, cannot be a coincidence but rather must be a natural and defining aspect of the pairing interactions in the cuprates. Such a short pairing length scale is not consistent with a conventional pairing mechanism that exchanges long-wavelength (low-$q$) phonons, or low-q spin fluctuations, but would likely admit a purely electronic mechanism [1] or perhaps one that exchanges zone-boundary bosons, such as the strong first-neighbor superexchange of antiferromagnetic interactions [22,23].

Although containing electrons moving in opposite directions, the antinodal pairs have an extremely short length scale that is extremely different from that of conventional superconducting pairs that extend over many lattice sites. Therefore, it allows a new type of question to be asked that would be meaningless for a system with large pairs: what is the location of the pair center relative to a specific lattice site, i.e. are the pairs centered on individual Cu atoms (**Fig. 4a**), bonds (**Fig. 4b**), or plaquette centers (**Fig. 4c**)? This is important since we should intuitively expect a different type of pairing interaction to be applicable depending upon which of the centering options of **Fig. 4** is realized, so this is addressing the most fundamental local nature of the pairing interaction. This question also gets at the very heart of the nature of the carriers in the cuprates – for example, even though spectroscopy indicates that the doped carriers (holes) predominantly go onto oxygen sites [24,25], strong theoretical arguments by Zhang and Rice indicate that the oxygen degrees of freedom can be projected out so that the effective carriers can be considered as Cu-centered [26] (a special type of Cu-O hybridized state). If the carriers are indeed Cu-centered Zhang-Rice singlets then we expect the pairs to be bond-centered (**Fig. 4b**), meaning they could go onto either horizontal or vertical bonds and among other things should lead to various nematic ordering tendencies similar to what has been observed in STM experiments [27]. On the other hand, if the carriers stay on the oxygen sites during the pairing [28] then one of **Fig. 4a** or **4c** would be expected. Future experiments, for example scanning Josephson tunneling microscopy experiments [29], might be able to resolve which of the centering options of **Fig. 4** the pairs choose. Further, the packing of starfish-shaped pairs onto the lattice would naturally lead to weak variations in charge density and pair density, potentially giving rise to charge density waves [30] and pair density waves [29] in the cuprates - a clear example of an intertwined order [31] between the superconducting and charge channels.

Because this pairing short length is so constant across the range of doping levels, it is plausible to consider it one of the defining characteristics of the pairing in the cuprates, perhaps more so than the strength of the pairing (superconducting gap) itself. Indeed, the strongly increasing pairing strength with underdoping, known for almost two decades [32], should perhaps no longer be considered to be a signal of an intrinsically stronger interaction in the underdoped regime, but rather as a means of fixing the antinodal pair size at this lattice-based value. A positive feedback on the pairing interactions, proposed recently [1], is one such mechanism that would allow or maybe force the pairing strength to grow until it sets the pair size at the optimal value, i.e. it enforces a type of speed-limit on the pairing enforced by the lattice parameters. In addition, such a short length scale at the antinode is very likely to be shorter than the mean distance between pairs, therefore, the antinodal pairs are unlikely to generate the long-range phase coherence of superconductivity. Rather, the phase coherence must be generated in the midzone or near-nodal directions where the length scale gets longer. This is an unusual situation where the portion of the zone that has the strongest pairing (and where the pairing is most likely generated) is not the part that is relevant for generating the actual superconductivity state of long-range phase coherence. The above indicates a possible mechanism for intertwined orders – we can affect the antinodal states with a charge density wave or a pseudogap, and as long as the rest of the k-states are minimally affected, then the long-range superconducting phase coherence can be maintained. Regardless of the mechanism of the pairing, our reconstruction of the shape and size of the pairs, including especially the universal antinodal length scale, will bring us much closer to understanding the various cuprate phenomenology as well as the microscopic interactions that drive the pairing in these materials.

**Methods**

Pb-doping renders BSCCO free of the superstructure effects which contaminate most ARPES data on BSCCO cuprates. Thus, we achieve data with very high resolution and low background, which is essential for self-energy analysis but is otherwise fully consistent with the large body of data taken by many groups over the years. ARPES experiments were performed at SSRL and HiSOR with energy resolution of ~10meV. In our measurements, we selected incident photon energy (24 eV) to separate the response of the bilayer-split antibonding band from that of the bonding band; results from former are shown here and those from the latter (which are qualitatively similar) are shown in Fig. S5 and Fig. S6, and discussed in the supplemental information S4. The angle-dependent pair sizes shown in Fig. 1 a and b are extracted from the ARPES data published in previous work [1].

We have performed high resolution ARPES on a series of superstructure-free bilayer $(Bi,Pb)_2Sr_2CaCu_2O_8$ (Bi2212) samples in both the overdoped and underdoped regimes and with $T_C$'s ranging from 69 to 91K, and with temperatures from 10K to 250K, though for the present work we focus on the low temperature superconducting-state spectra. A key enabler of the present work was the recent development of a new two-dimensional fitting procedure for the ARPES data [1] that no longer must focus on the one-dimensional MDCs (Momentum Distribution Curve) or EDCs (Energy Distribution Curves), but instead treats these simultaneously and on an equal footing. Our analysis assumes (successfully so far) that spectral broadening and band renormalizations (the effects of electronic correlations) can be treated within the conventional many-body language of Green's functions and electronic self-energies, though we do not restrict the system to have self-energies that are small compared to the energy of the states, i.e. there is no requirement that we reside in the quasiparticle or Fermi-liquid frameworks. With this we are able (successfully so far) to describe all modifications to the spectra due to the onset of superconductivity within the framework of the conventional Nambu-Gor'kov formalism. Another new aspect of this procedure is that the real and imaginary parts of the extracted electronic self-energies $\Sigma'(\mathbf{k},\omega,T)$ and $\Sigma''(\mathbf{k},\omega,T)$ are intrinsically constrained to be causal (obey the Kramers-Kronig relations), as opposed to previous efforts in which the Kramers-Kronig relations were taken in one direction (e.g. taken from $\Sigma''$ to obtain information about $\Sigma'$, or vice-versa). This treatment allows us to extract, naturally and simultaneously, the superconducting gaps, the coherence factors for particle-hole mixing (which determine the Bogoliubov quasiparticles), and both real and imaginary parts, $\Sigma'$ and $\Sigma''$, of the complex self-energy. All of these quantities were obtained at many momentum points throughout the Brillouin zone and for many temperatures across the pairing and superconducting regimes.

**Acknowledgements**

We thank Bruce Normand and Rahul Nandkishore for very useful conversations, Mark Golden for putting us in touch with Y. Huang for sample growth, and D. H. Lu and M. Hashimoto for assistance at SSRL. Funding for this research was provided by DOE Grant No. DE-FG02-03ER46066 and by the Center for Experiments on Quantum Materials. SSRL is operated by the Office of Basic Energy Sciences of the US Department of Energy and HiSOR is supported by the Japanese Ministry for Science. Work at Brookhaven was supported by the Office of Basic Energy Sciences (BES), Division of Materials Science and Engineering, U.S. Department of Energy (DOE), through Contract No. DE-sc0012704. R. D. Z. and J. S. were supported by the Center for Emergent Superconductivity, an Energy Frontier Research Center funded by BES.

*Correspondence and requests for materials should be addressed to D.S.D. (dessau@colorado.edu), H.L. (haoxiang.li@colorado.edu)

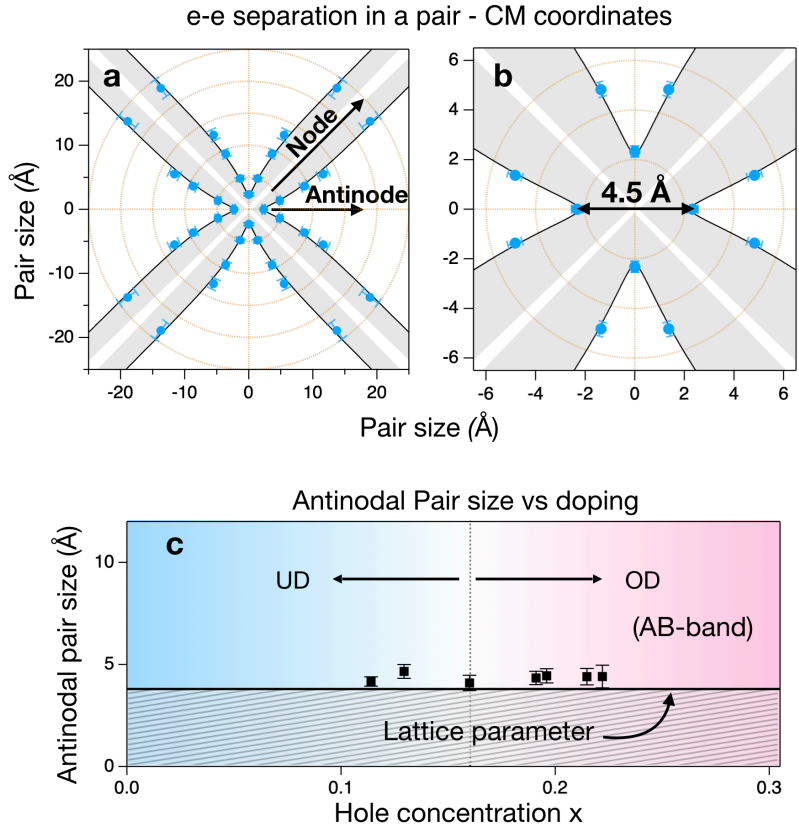

**Figure 1 | The angular dependence of the Cooper pair length scales as extracted from ARPES data. a.** Pair size (expectation value of electron-electron separation using center-of-mass (CM) coordinates) as a function of direction measured from a Tc=85K under doped $Bi_2Sr_2CaCu_2O_8$ sample (UD85K) at 15K. The pairs extend to >50Å near the nodal direction, with drastically decreasing size when moving towards the antinode. **b.** A zoom in to the central region of a pair. An ultrashort length scale of 4.5Å is seen for the antinodal pairs – a length on the order of the lattice parameter. **c.** The ultrashort antinodal length scale is maintained for all doping levels studied indicating this is likely a universal property of the Cooper pairs. Raw ARPES data and relevant parameters throughout the Brillouin zone are obtained from ref [1]. See text and supplemental information for details of how the pair size is obtained as well as the impact of the bilayer splitting (the present data is only from the antibonding or AB band). The error bars of the pair size are based on the uncertainty of the parameters that are used for the calculation.

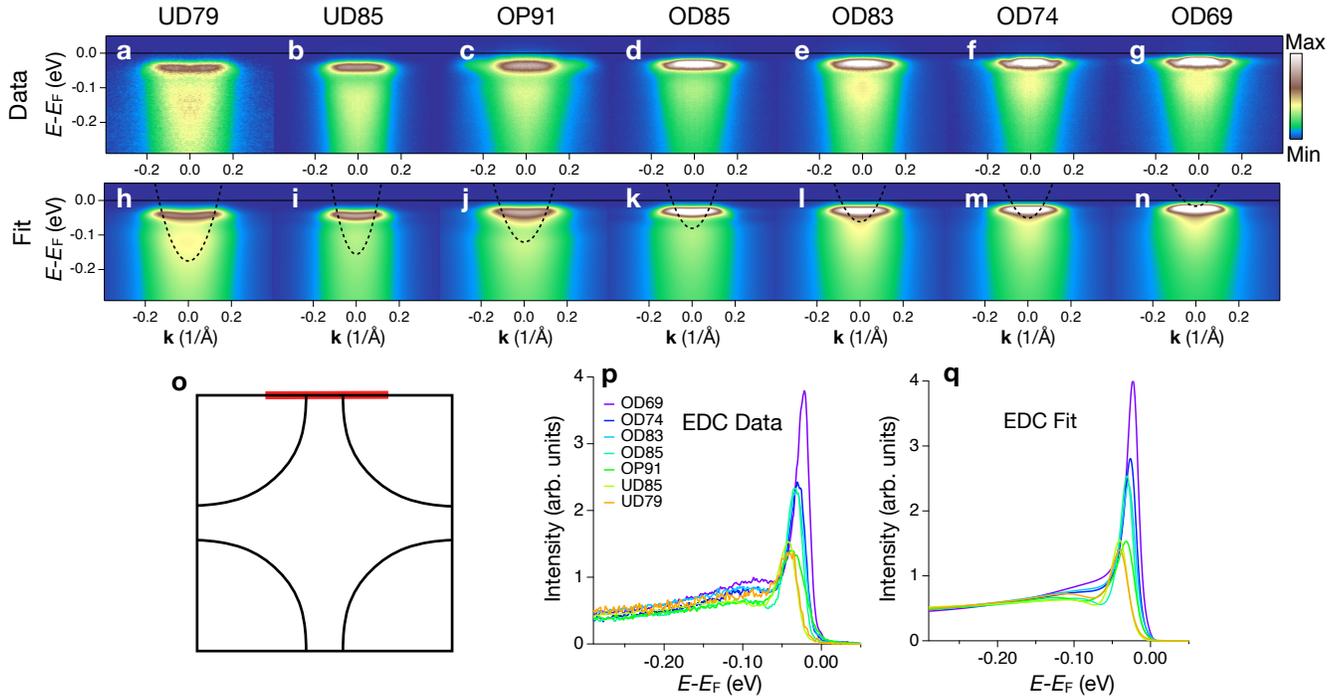

**Figure 2 | Antinodal spectra and fits over a wide doping range at temperatures well below $T$c. a-g.** Experimental spectra, with photon energy chosen so as to emphasize the antibonding band. **h-n**. The corresponding 2D fitting results plotted together with the extracted bare bands (black dashed curves). Intensity scales between doping levels are constrained by the theoretical fits and so are the measured spectra. **o.** Fermi surface plot including the location of the cuts of panels **a-g** as shown by the red line. **p-q.** $k_F$ EDCs and fits taken from the top panels. UD79 means under-doped sample with Tc of 79K, OP91 means optimal doped sample with Tc of 91K, OD85 means over-doped sample with Tc of 85K, etc.

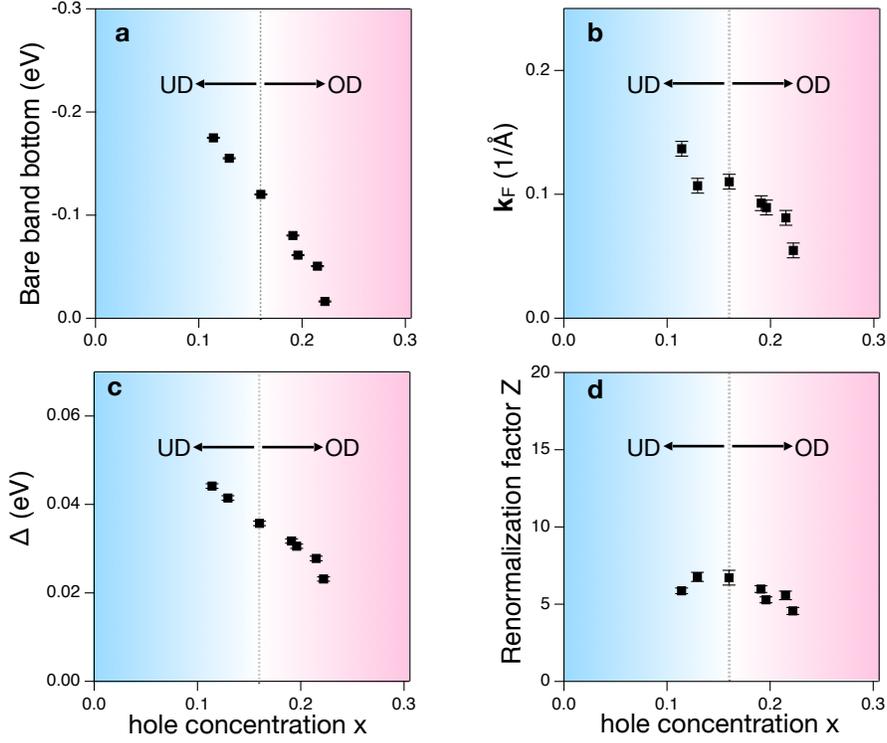

**Figure 3 | The key parameters vs doping extracted from the 2D fitting to the antinodal spectra. a-b.** Two parameters that we used to construct the bare band for the antinodal cut— the bare band bottom (panel **a**), and the momentum at the Fermi level $k_F$ (panel **b**). Both of these parameters display an expected decreasing trend as moving towards the over doped side. **c.** the superconducting gap size. It follows a linear trend with the doping, which is consistent with the previous findings in the literature [33]. **d.** Renormalization factor Z, extracted from the real part of the self-energy at $E_F$ (see supplementary S5 and Fig. S7 for detailed description.). The error bars in panel **a** and panel **c** include the uncertainty of the $E_F$ (±0.5 meV), and the 3σ return from fitting. The error bars in panel **b** contain the momentum resolution (±0.006 1/Å), and the 3σ return from fitting. The error bars in panel **d** reflect the standard deviation of the Z value extracted from different energy range near the Fermi level (within the superconducting gap energy scale).

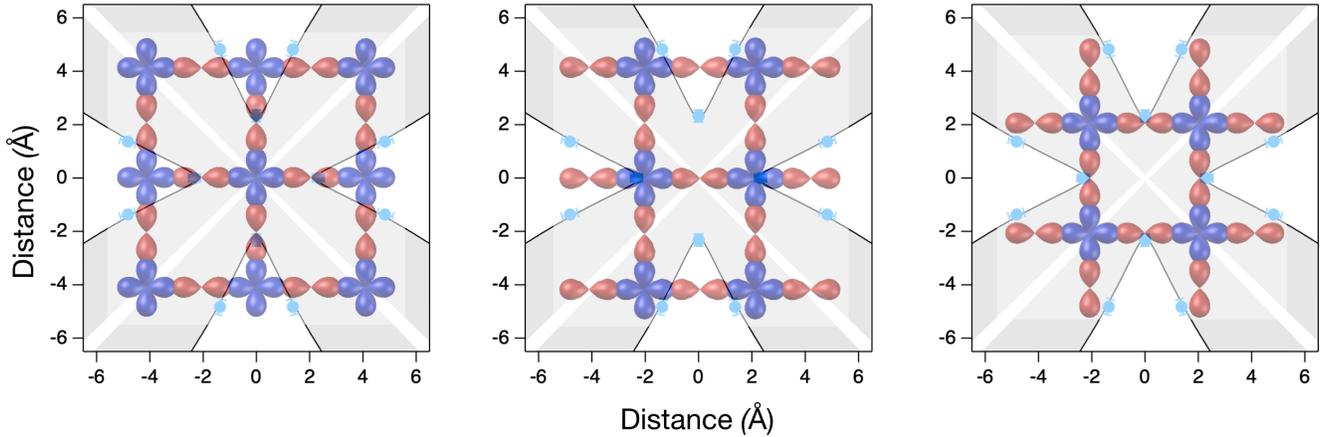

**Figure 4 | Various centering options for the pairs as drawn using CM coordinates.** Because the pair length scales are so short it is reasonable to expect them to be centered at specific locations within the Cu-O lattice, with the three main options shown here: **a**. A Cu-centered pair. **b**. A horizontal bond-centered pair. Vertical-bond centered pairs should be energetically degenerate with this configuration. **c**. A plaquette-centered pair. Cu atoms are shown as brown $d_{x^2-y^2}$ orbitals, oxygen as beige p orbitals.

# Supplementary Information

## S1. Weighted average of Pippard coherence length

For a BCS superconductor, the coherence length is given by

$$\xi_{CL}(\theta) = \frac{\hbar v_F}{\pi \Delta} \tag{S1}$$

In the presence of strong correlations, the Fermi velocity $v_F$ is given by $v_F = v_{BB}/Z$ with $v_{BB}$ the bare band velocity and $Z$ the renormalization factor. Thus, $v_F$ is an effective quantity whose strong suppression is contained in the renormalization parameter $Z$ which may also be considered as an enhancement of the single-particle mass. **Fig. S1** shows the coherence length (panel **a**) of sample UD85K calculated according to Eqn. S1, using the bare bands $\varepsilon_{0\mathbf{k}}$, $\Delta$, $\Sigma''$, and $Z$ that are extracted from the 2D fitting. To get a weighting for the coherence length at different Fermi surface angles, we consider the condensate density of Cooper pairs $N_c = \Delta(\mathbf{k})^2/E_\mathbf{k}^2$ [1,2], where $\Delta$ is the superconducting gap, $E_\mathbf{k} = \sqrt{\varepsilon_\mathbf{k}^2 + \Delta(\mathbf{k})^2}$ with $\varepsilon_\mathbf{k}$ is the non-gapped dispersion (with renormalization). The integrated condensate density (normalized to maximum) varies along Fermi surface angle as shown in **Fig. S1b**. The condensate density is 0 along the nodal direction and gradually increases until it is maximal at the antinode. We can then calculate the average coherence length weighted by the condensate density to be 13Å for a UD85 sample and 17Å for a OD74 sample. We plot these values in **Fig. S2** (red points) together with values from the literature extracted from Hc2 measurements from reference [3] (blue dots). Converting the Hc2 data to coherence length using the formula $\xi_{CL} = \sqrt{\Phi_0/2\pi H_{c2}}$ gives the dashed red line of **Fig. S2**, which within the error bar (up to a factor of two in the $H_{c2}$ measurements) compares reasonably well to the ARPES-extracted coherence lengths.

## S2. The functional form of pair size

The general expression of the characteristic size of the Cooper pairs is:

$$\xi_{pair} = \sqrt{\frac{\int dr |\psi(r)|^2 r^2}{\int dr |\psi(r)|^2}} = \sqrt{\frac{\int dk |\partial_k \psi(k)|^2}{\int dk |\psi(k)|^2}} \tag{S2}$$

where $\psi(r)$ is the Cooper-pair wave function in real space, $\partial_k$ is the gradient operator in momentum space (**k** space), and $\psi(\mathbf{k})$ is the Fourier transform of $\psi(r)$ to **k** space [4]. We utilize the standard result $\psi(\mathbf{k})=\Delta(\mathbf{k})/E_\mathbf{k}$, with $E_\mathbf{k} = \sqrt{\varepsilon_\mathbf{k}^2 + \Delta(\mathbf{k})^2}$ [4], where $\varepsilon_\mathbf{k}$ is the non-gapped dispersion that includes the renormalization effect due to the many-body interactions. $\varepsilon_{0\mathbf{k}}$ is the bare band dispersion. For the antinodal cuts as shown in Fig. 2, we can approximate the bare band as a parabola, i.e. $\varepsilon_{0k} = -\frac{E_{BB}}{k_F^2}\omega^2 + E_{BB}$ with parameters of the Fermi momentum $k_F$ and bare band bottom $E_{BB}$, where the $k_F$ and $E_{BB}$ of the antinodal cut of various doping are displayed in Fig. 3. For the angle dependent result (Fig. 1a and b), we use a tight binding band structure that based on the bare band dispersions extracted from the different ARPES cuts through the Brillouin zone (shown in ref. [7]). We evaluate Eqn. S2 for every angle $\theta$, discussing the denominator and numerator separately. The denominator of the pair size as a function of angle is:

$$D_\theta = \sqrt{\int dk_\perp \frac{\Delta_\theta^2}{(\varepsilon_{0k}/Z_\theta)^2 + \Delta_\theta^2}} \tag{S3}$$

where the **k** integral is perpendicular to the Fermi surface at an angle $\theta$, the integral range is within the first Brillouin zone, and $\varepsilon_{0\mathbf{k}}$, $\Delta_\theta$, and $Z_\theta$ are the corresponding bare band, superconducting gap and renormalization factor for the ARPES cut at certain Fermi surface angles (the antinodal cut is shown in Fig. 2), and $E_{\mathbf{k}} = \sqrt{\varepsilon_{\mathbf{k}}^2 + \Delta_\theta^2} = \sqrt{(\varepsilon_{0k}/Z_\theta)^2 + \Delta_\theta^2}$. The numerator of the pair size functional form is:

$$N_\theta = \frac{1}{Z^2}\sqrt{\int dk_\perp \frac{\varepsilon_{0k}^2 \Delta_k^2}{E_k^6}(\partial_k \varepsilon_{0k})^2 + \frac{\varepsilon_{0k}^4 \Delta_k^2}{E_k^6}(\partial_k \Delta_k)^2 - \frac{2\varepsilon_{0k}^3 \Delta_k}{E_k^6}(\partial_k \varepsilon_{0k})(\partial_k \Delta_k)} \quad (S4)$$

with the same integral condition with the numerator described above. Here $\partial_k = \partial_\perp + \partial_\parallel$ where $\partial_\perp$ is the derivative normal to the Fermi surface and $\partial_\parallel$ is the derivative tangential to the Fermi surface. However, as the variation of $\Delta(\mathbf{k})$ (within 40 meV) over the whole Brillouin zone is much smaller than the variation of $\varepsilon_{0\mathbf{k}}$ (~0.5 eV), thus $\partial_k \Delta \ll \partial_k \varepsilon_{0k}$ and the first term in the numerator dominates. On the other hand, $\partial_\parallel \varepsilon_{0k} \ll \partial_\perp \varepsilon_{0k}$ so the dominant term (subscript D for dominant added) in the numerator can be approximated as

$$N_\theta \approx N_{\theta D} = \frac{1}{Z^2}\sqrt{\int dk_\perp \frac{\varepsilon_{0k}^2 \Delta_\theta^2}{E_k^6}(\partial_\perp \varepsilon_{0k})^2} \quad (S5)$$

To further justify the approximation of $N_\theta \approx N_{\theta D}$, **Fig. S3** shows a simulation result of the percentage variation of $N_D$ to $N$ as $(1 - \frac{N_{\theta D}}{N_\theta}) \times 100\%$ throughout the Brillouin zone. The largest variation is only about 3%, which is much smaller than our systematic error for the pair size.

### S3. Cooper pair size and coherence length

In the BCS limit (band bottom energy much larger than the superconducting gap size $E_{BB} \ll \Delta$), the Cooper pair size $\xi_{pair}$ is quite similar to the coherence length $\xi_{pair} \approx 1.11 \xi_{CL}$ as can be obtained by a direct calculation using Eqn. S5. However, in cuprate superconductors, when moving towards the antinode, $\Delta$ increases, and the band bottom decreases, the ratio of pair size to coherence length deviates from the BCS limit (**Fig. S4a**). And in fact, at the antinodal region, samples of all dopings that we measured show a strong deviation from the BCS limit (**Fig. S4b**) due to the shallow band bottom that is comparable or even smaller (OD69) than the superconducting gap.

### S4. Bilayer splitting and Bonding and Antibonding bands.

Due to the fact that there are two $CuO_2$ planes per unit cell (a bilayer) with some coupling between them, the bands split into a lower or bonding band (B) and an upper or antibonding band (AB) [5,6]. By utilizing a photon energy of 24 eV the AB band has a dominant matrix element (experimental intensity) and there is almost no contamination from the B band. This is the photon energy that was utilized for all data shown in the main text. Utilizing the photon energy of 20 eV both the B and AB bands are visible, as shown in **Fig. S5c** for the OD69 sample. The very clean separation of the two bands for this sample and this photon energy allows us to do a proper 2D fitting and extract all relevant parameters, as summarized by the green data points in **Fig. S6**. While a similar effect is observed for other dopings, the distinction between the two bands is less clear and so an accurate fitting is presently more difficult. As expected the bare band bottom and $k_F$ are very different for the B band, while the extracted gap values are essentially identical. The renormalization parameter Z is perhaps slightly smaller than for the AB band, though the difference is minimal. The B band pair size is however dramatically larger than the AB band pair size.

### S5. Extraction of the parameter Z from the slope of $\Sigma'$ vs $\omega$.

It is perhaps surprising that the renormalization parameter *Z* has only a weak doping dependence (**Fig 3d**). The way this is extracted from our data is shown in **Fig. S7**. First $\Sigma''(\omega)$ and $\Sigma'(\omega)$ were extracted from the low temperature antinodal data from each sample utilizing the recently-developed 2-dimensional fitting technology [7]. The value Z is defined $Z = 1 + |\partial\Sigma'/\partial\omega|_{\omega\to 0}$, i.e. it is related to the slope of $\Sigma'$ vs $\omega$ at $E_F$. This slope is highlighted by the black dashed line in **Fig. S7a**. It is observed that the low energy slope is approximately independent of doping for this low T antinodal data, explaining the approximately constant value of Z shown in **Fig 3d**.

**S6. Pair size, pair radius, and the Center of Momentum frame.**

Equations S3 and S6 show that the pair size is calculated as the root-mean-square expectation value of the radius $r$. Utilizing the experimentally determined pair wavefunction, a direct calculation returns values that range from $r$=4.5 Å at the antinode. This value of $r$ is the inter-electron separation within a pair, i.e. it gives the separation of one of the electrons from the other electron, just as it would give the separation of an electron from the nucleus if a one-electron hydrogenic wavefunction was utilized in Eqn. S2. In contrast to a H atom with a heavy nucleus that has the center of momentum (CM) at the nucleus, the two electrons in the Cooper pair have equal mass so a reduced mass scheme with the CM midway between the two electrons is more intuitive for visualization. This is the scheme utilized for Fig. 1 and Fig. 4, i.e. the expectation values of the pair are observed to go between $\pm r/2$.

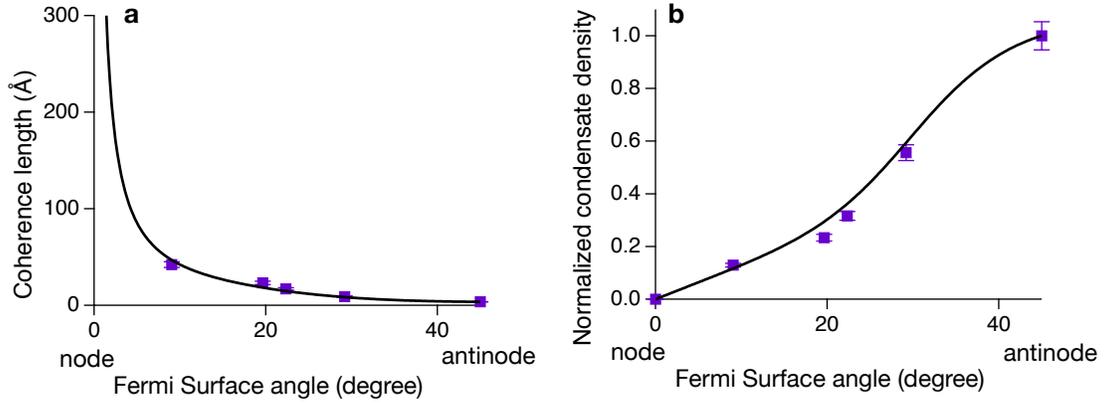

**Figure S1 | Coherence length of the sample UD85K at 15K. a.** Coherence length vs Fermi surface angle, calculated using Eqn. S1 with all the parameter extracted from 2D fitting method (purple square). The corresponding ARPES data and the fit results are presented in ref. [7]. The black curve is an interpolation of the data points. **b**. The calculated condensate density of Cooper pairs vs. Fermi surface angle using the same set of parameters, and normalized to the value at the antinode. Again, the black curve is an interpolation of the data points. The expression of the condensate density is presented in supplementary information S1. The error bars in the figures are based on the uncertainty of the parameters extracted from the 2D fitting. Using the interpolated curves shown in the figure, we can then calculate the properly averaged coherence lengths.

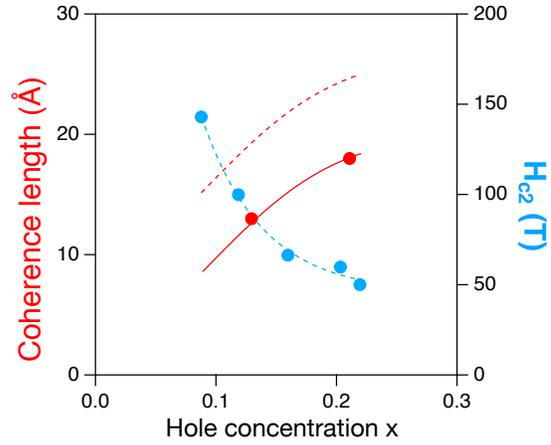

**Figure S2**. The weighted average coherence length extracted from ARPES across the Brillouin zone with two doping levels (red dots, left axis) is compared with $H_{c2}$ data of Bi2212 extracted from Wang et al [**3**] (blue, right axis) and converted to coherence length (red dashed line, left axis). The red solid line is a constant shift of the red dashed line along the Y axis.

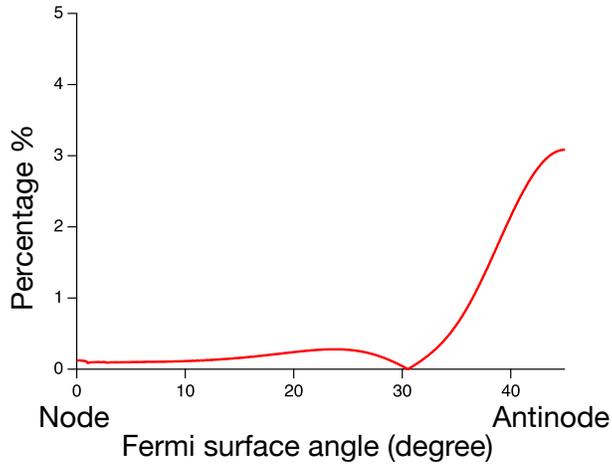

**Figure S3 | Simulation result of the percentage variation of the approximate numerator terms to full term in the pair size functional form.** The maximum variation is only ~3%.

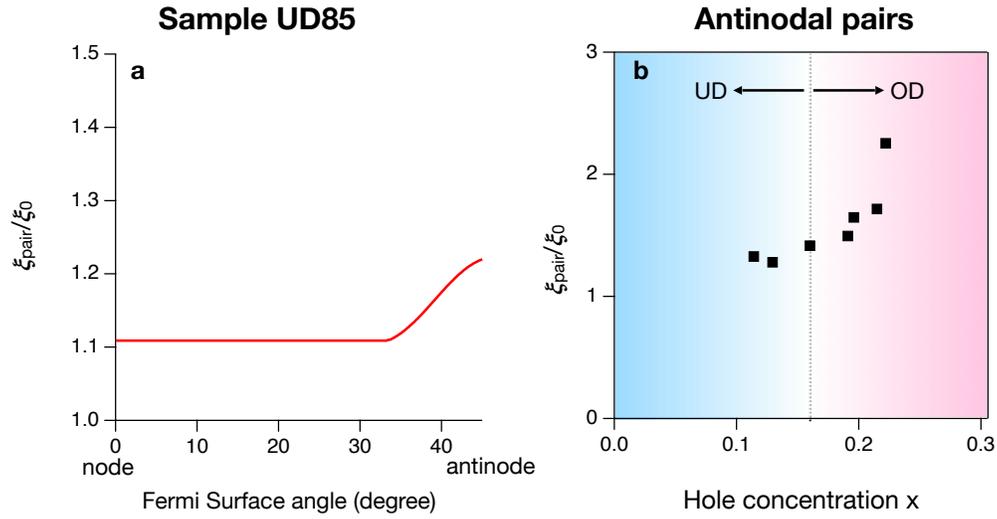

**Figure S4 | Ratio of Cooper pair size and coherence length. a.** Ratio of Cooper pair size and coherence length vs Fermi surface angle of sample UD85K at temperature 15K. The majority of the Fermi surface is within the BCS limit ($\Delta \ll E_{BB}$) with the ratio of ~1.11. The ratio rises up towards antinode as the gap size increases and the band bottom energy decreases. **b.** Ratio of the antinodal Cooper pair size and coherence length vs doping. The upturn in the overdoped region is due to the rapidly decreasing energy of the band bottom (Fig. 3 in main text).

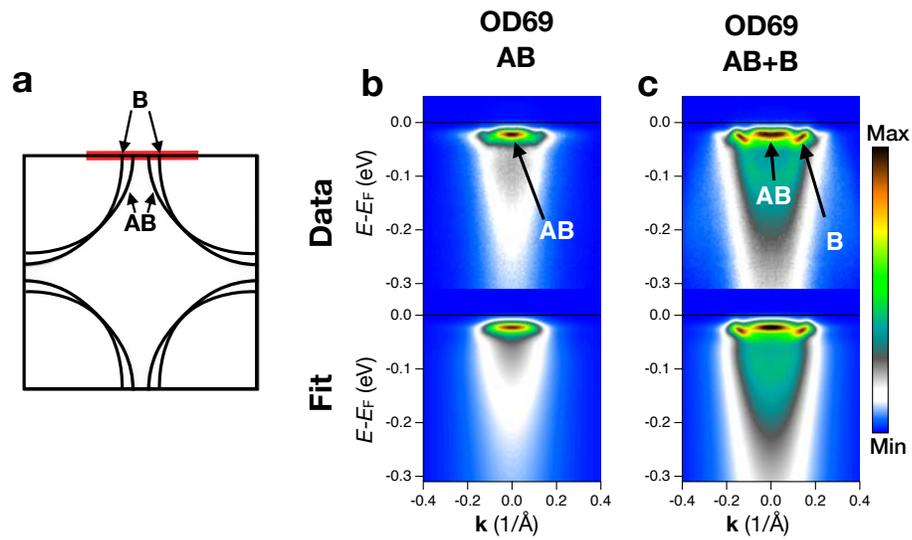

**Figure S5 | Bilayer split bands.** Antibonding=AB, Bonding=B. Antinodal ARPES from the OD69 sample taken with different photon energies so as to emphasize the AB band only (panel **b**) or both the AB and B bands (panel **c**). Panel **a** shows the experimental Fermi surfaces for both the AB and B bands. hv=24 eV was utilized to access the AB band while 20 eV photon was used to emphasize both of them.

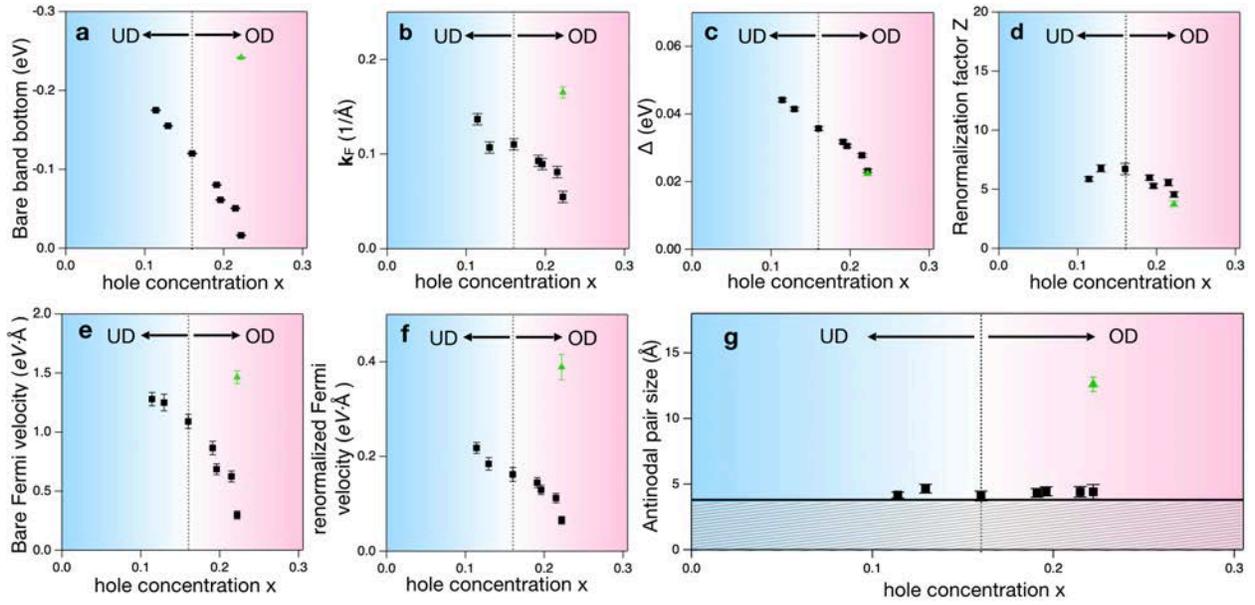

**Figure S6. Comparison of bonding band parameters (green triangle) vs. antibonding (black square).** The bonding band pair sizes are much larger than the antibonding pairs. The bare Fermi velocity ($v_B$) is calculated from the extracted bare band dispersion shown in Fig. 2h-n of the main text. The renormalized Fermi velocity ($v_{re}$) is given by $v_{re}=v_B/Z$, where $Z$ is the renormalization factor.

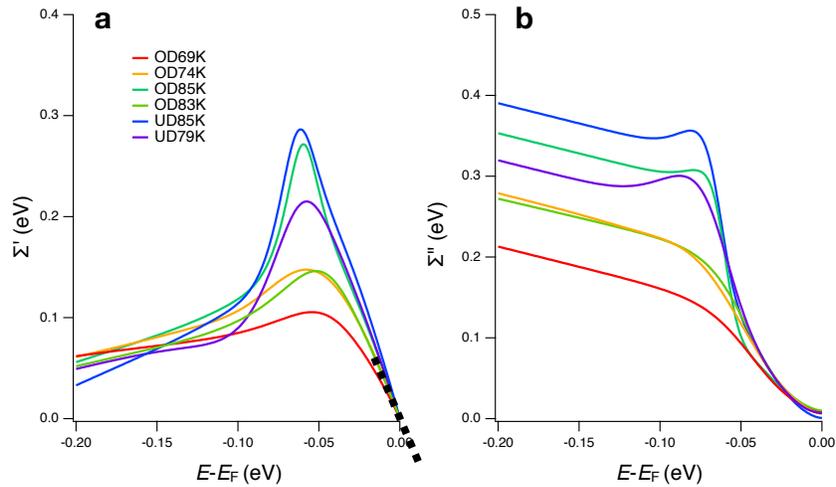

**Figure S7 | Self energy of the antinodal cut with different dopings at temperature well below Tc.** The real part (panel **a**) and the imaginary part (panel **b**) of the self-energy extracted from 2D fitting results shown in Fig. 2. The slope of Σ′ near $E_F$ (black dashed line) is approximately independent of doping, and is the reason that Z is roughly constant with doping.

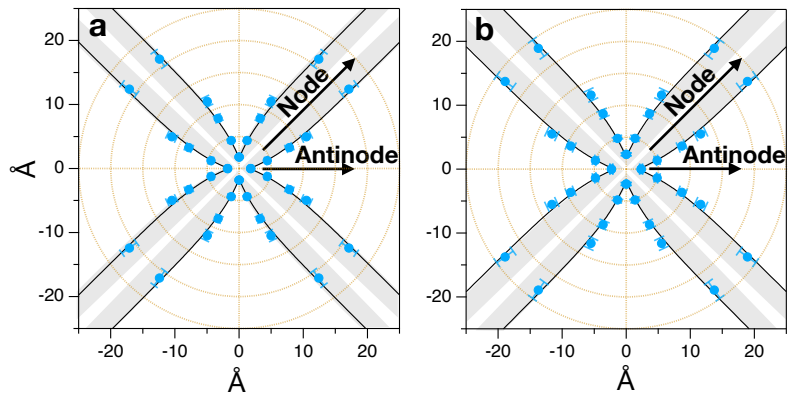

**Figure S8 | Comparison of coherence length and pair size.** Polar plot of coherence length (panel **a**) with the value divided by an extra factor of 2 in order to compare with the pair size plot in the center of mass coordinate (panel **b**, the same plot as Fig. 1a).